\documentclass[12pt]{article} % For LaTeX2e

\usepackage[colorlinks=true]{hyperref}
\usepackage[margin=1.in]{geometry}
\usepackage{graphicx}
\usepackage{amsmath}
\usepackage{amssymb}
\usepackage{booktabs}
\usepackage{caption}
\usepackage{subcaption}
\usepackage{verbatim}
\usepackage{float}
\usepackage[english]{babel}
\usepackage{braket}
\usepackage{xy}
\usepackage{longtable}
\usepackage{tabularx}

\title{Evaluating a Digital Speech Therapy App for Stuttering: A Pilot Validation Study}

\author{Urvisha Shethia \\ \small{Kindering Healthcare}
        \and Vedali Inamdar \\ \small{Vaikhari Speech Therapy}
        \and Viraj Kulkarni \\ \small{Iyaso}}

\begin{document}
\hyphenpenalty=1000
\date{}

\maketitle

\begin{abstract}
\noindent Stuttering is a clinical speech disorder that disrupts fluency and leads to significant psychological and social challenges. This study evaluates the effectiveness of Eloquent, a digital speech therapy app for stuttering, by analyzing pre-therapy and post-therapy speech samples using the Stuttering Severity Index-4 (SSI-4) and the S24 communication and attitude scale. Results showed a 52.7\% reduction in overall SSI-4 scores, with marked improvements in reading (45\%), speaking (46\%), duration (57\%), and physical concomitants (63\%) scores. Over 75\% of participants improved by at least one severity category. S24 scores decreased by 33.5\%, indicating more positive self-perceptions of speech and reduced avoidance. These findings highlight the potential of structured, technology-driven speech therapy interventions to deliver measurable improvements in stuttering severity and communication confidence.
\end{abstract}

\section{Introduction}
Stuttering is a speech disorder characterized by involuntary disruptions in fluency, often leading to communication difficulties, social anxiety, and diminished self-confidence \cite{world1992icd}. It affects approximately 1\% of the global population, with varying degrees of severity \cite{yairi1992longitudinal}. While traditional speech therapy approaches such as fluency shaping and stuttering modification techniques can improve fluency, they often require consistent access to trained speech-language pathologists (SLPs), making them costly and inaccessible for many individuals who stutter. Digital interventions have emerged as a promising alternative, offering structured, self-guided exercises that allow individuals to practice fluency techniques in a scalable and convenient manner. Eloquent is a speech therapy app designed to provide guided practice and confidence-building exercises for people who stutter. It integrates evidence-based fluency-shaping and stuttering modification techniques such as light contacts, gentle onsets, preparatory sets, pull-outs, cancellations, etc. into an accessible digital platform.
\newline \newline
This pilot validation study evaluates the effectiveness of Eloquent in reducing stuttering severity and enhancing speaking confidence. Using the Stuttering Severity Instrument-4 (SSI-4) and the Erickson S24 scale, we assess pre-therapy and post-therapy changes in fluency and self-perceived communication attitudes. Participants completed a 15-session intervention, with progress tracked through video-based speech assessments. Blinded scoring by a qualified SLP ensured objective evaluation. By quantifying the impact of Eloquent on both fluency and confidence, this study contributes to the growing body of research on digital speech therapy. The findings will help determine the feasibility of digital interventions as an accessible alternative to traditional therapy and inform the development of evidence-based tools for people who stutter.

\section{Background}
While numerous clinical validation studies and randomized controlled trials have assessed interventions for children and adolescents who stutter, this study focuses exclusively on adults. Various therapeutic approaches for adult stuttering have been explored for their effectiveness \cite{brignell2020systematic, baxter2016non}, with many interventions combining speech restructuring techniques and psychological components.
\newline \newline
The Camperdown Program, which employs prolonged speech patterns, has demonstrated significant fluency improvements, sustained over a 12-month follow-up in a multicenter trial. However, individual outcomes varied in terms of fluency gains and long-term maintenance \cite{o2018camperdown}. Carey et al. \cite{carey2010randomized} conducted a randomized controlled trial comparing telehealth and face-to-face delivery of the Camperdown Program and found no significant differences in outcomes, supporting the viability of remote therapy. Similarly, Cream et al. \cite{cream2010randomized} reported that video self-modeling post-treatment was associated with improvements in long-term, self-reported fluency outcomes. Other studies have compared different speech restructuring techniques. Ingham et al. \cite{ingham2015efficacy} examined prolonged speech versus modified phonation intervals, while Menzies et al. \cite{menzies2008experimental} evaluated the effectiveness of prolonged speech with and without cognitive behavioral therapy (CBT). Telehealth methodologies have also been extensively reviewed. Lowe et al. \cite{lowe2014review} found that 82\% of studies reported significant clinical outcomes with speech therapy delivered via telephone, webcams, or self-guided internet programs. Brignell et al. \cite{brignell2020systematic} provided a comprehensive review of speech therapy treatments, and McGill et al. \cite{mcgill2019telepractice} focused specifically on interventions delivered via telepractice.
\newline \newline
Collectively, these studies confirm the general effectiveness of speech therapy in reducing stuttering, emphasizing the need for personalized treatment plans and adjunctive therapies. The growing body of evidence on telehealth and digital interventions underscores the potential of technology-driven approaches to improve accessibility and long-term treatment outcomes for adults who stutter.

\section{Intervention Overview}
\begin{figure}[ht]
\centering
\includegraphics[width=\linewidth]{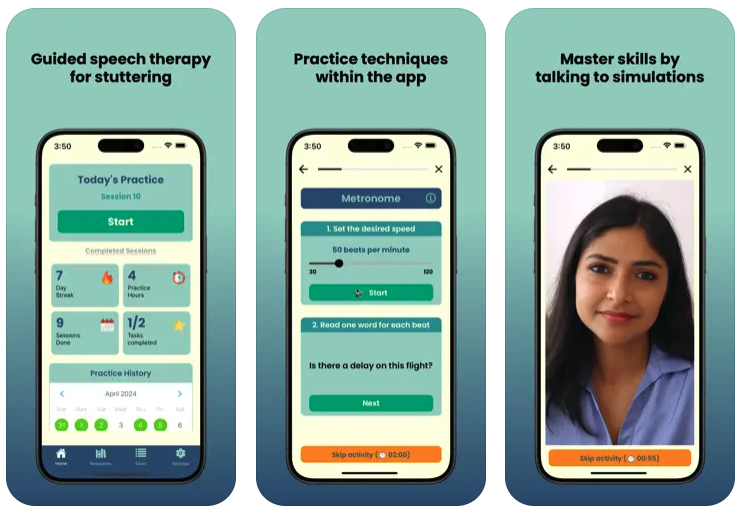}
\caption{Screenshots from Eloquent: The home screen showing progress and metrics (left); Metronome tool to practice rhythmic speech cueing (center); Virtual avatar that simulates the experience of speaking to a real person (right)}
\label{fig:eloquent}
\end{figure}

Eloquent is a self-guided stuttering therapy application available on Android and iOS devices. The program consists of structured sessions, each lasting approximately 10 minutes, which introduce users to a range of evidence-based techniques designed to improve speech fluency. These techniques include rhythmic speech cueing, light contacts, gentle onsets, preparatory sets, pull-outs, cancellations, etc. Beyond fluency shaping and stuttering modification, Eloquent addresses avoidance behaviors and emphasizes anxiety regulation in social speaking situations. The app incorporates targeted exercises such as voluntary stuttering and 4x4 breathing to help users develop confidence and control in real-world interactions.
\newline \newline
Simulations provided through the application allow users to practice speaking with virtual avatars. This serves as an intermediate step between solo practice and real-life conversations, enabling users to build fluency in a low-pressure environment before applying their skills in everyday situations. To ensure the transfer of learned techniques into daily life, the app assigns real-world speaking tasks that users complete and log within the platform. These tasks include challenges such as \textit{talk to a friend about your stuttering}, \textit{use gentle onsets in a conversation}, \textit{practice 4x4 breathing before an anxiety-inducing conversation}, etc. By reinforcing skill application beyond the digital environment, Eloquent bridges the gap between therapy and real-world communication.
\newline \newline
To enhance user engagement and adherence to therapy, Eloquent integrates gamification elements such as notifications, progress tracking, motivational reminders, and streaks. Users unlock new tools and exercises as they complete program sessions and real-world speaking challenges, providing the motivation to remain consistent with their practice. A personalized dashboard visualizes improvement over time, reinforcing positive reinforcement and commitment to the intervention. By incorporating these engagement mechanics, Eloquent transforms stuttering practice into a rewarding experience, thereby increasing adherence and maximizing outcomes. Figure \ref{fig:eloquent} shows screenshots from the Eloquent mobile application.

\section{Study Method}
Data collection for this study followed a structured and systematic approach, ensuring that all participants underwent the same procedures and assessments throughout the intervention period.

\subsection{Participant Registration}
Interested users of the Eloquent app were invited to participate in the study via a message displayed on the app's home screen. To begin the enrollment process, participants were required to complete a Google Form that collected their personal information and confirmed their eligibility for the study. To be deemed eligible, participants had to meet three criteria: (1) a commitment to using the Eloquent app for 10 minutes daily over 15 days, (2) willingness to participate in two video calls via Google Meet or Zoom, and (3) consent to undergo speech assessments during these sessions. Eligible participants received a welcome email with a scheduling link and instructions to schedule their first video assessment.

\subsection{Pre-Therapy Video Assessment}
A pre-therapy assessment was conducted via a video call using Google Meet or Zoom. In instances where Internet bandwidth was insufficient to record high-quality video, participants were asked to use their phone cameras to record a selfie video while remaining present on the video call. During each video session, participants were first asked to introduce themselves and engage in a spontaneous two-minute conversation on a topic of their choice, such as describing their city or any other subject they preferred. This was followed by reading aloud the Rainbow Passage, a commonly used passage in speech assessments.

\subsection{Intervention}
Participants who completed the pre-therapy assessment were granted free access to the Eloquent app and were asked to complete one session per day. Post-therapy assessments were conducted after participants completed 15 sessions. To accommodate personal commitments, a five-day grace period was provided, allowing participants up to 20 days to finish the intervention.

\subsection{Post-Therapy Video Assessment}
A post-therapy assessment was conducted via a video call following the same procedure as the pre-therapy assessment. This assessment was available only to participants who completed all 15 program sessions within the 20-day period.

\subsection{Self-Reported Questionnaire}
In addition to the SSI-4 assessments, the participants' speaking confidence was measured using the Erickson S24 scale \cite{andrews1974stuttering}. The S24 questionnaire was administered as a self-reported survey through the Eloquent app at the beginning and after Session 15 of the guided therapy program. Participants completed the S24 questionnaire independently within the app, and the results were automatically scored by the app based on the Erickson S24 scale.

\subsection{Follow-ups and Dropouts}
A total of 33 participants submitted the registration form, after which enrollment was closed. Of these, 27 participants scheduled their first video call, and 23 attended and completed the pre-therapy assessment. These participants were then granted free access to the Eloquent app and instructed to complete one session daily. A 20-day window, including a 5-day grace period, was provided for participants to complete 15 sessions. App usage logs were monitored to verify session completion before scheduling the post-therapy assessment. Of the 23 participants who completed the pre-therapy assessment, 17 successfully completed all 15 sessions and participated in the post-therapy assessment.

\subsection{Collection of Consent}
Participants provided consent via the registration form by agreeing to the statement: \textit{The data gathered during this program will be anonymized and used for clinical research purposes. Your personal information and contact details will remain confidential and will only be used for contacting you about this program.} Additionally, further consent was secured via email, informing participants of the following: \textit{(1) Your data will be used for research purposes including determination of your speech scores across different assessments; (2) Your data may be shared with clinical and technology experts, consultants, and team members directly involved in the research; (3) Your data will NOT be shared with any third party not directly involved in the research.}

\section{Scoring}
Both pre-therapy and post-therapy video recordings were collected and batched together for evaluation. A qualified speech-language pathologist (SLP), who was blinded to the participants' names, dates, and whether the recordings were pre-therapy or post-therapy, reviewed and scored each video using the Stuttering Severity Index-4 (SSI-4). The SSI-4 \cite{rileyssi} is a widely used measure of stuttering severity, which takes into account the frequency of stuttering, duration of stuttering events, and severity of the physical concomitants accompanying stuttering. The final SSI-4 score is the sum of the frequency score, duration score, and the physical concomitants score.
\newline \newline
We followed the standard SSI-4 guidelines to calculate these component scores. The frequency score was determined by calculating the percentage of syllables stuttered during the recorded speaking and reading speech samples, converting them separately into the speaking task score and reading task score using the standardized SSI-4 scoring chart, and then adding these two task scores together. The duration score was derived by averaging the three longest instances of stuttering within the speech sample and converting this average duration into a scaled score. The physical concomitants score was assessed based on four categories of observable secondary behaviors accompanying stuttering namely distracting sounds, facial grimaces, head movements, and movement of the extremities. These behaviors were each rated on a severity scale ranging from 0 (no noticeable physical concomitants) to 5 (severe and painful looking), and the physical concomitant score was the sum of them all. The final overall SSI-4 score was obtained by summing the frequency score, duration score, and the physical concomitant score, which categorized the participant’s stuttering severity into one of five levels: very mild, mild, moderate, severe, or very severe. This comprehensive scoring approach allowed for an objective comparison between pre-therapy and post-therapy assessments, providing a quantifiable measure of any improvements in stuttering severity following the intervention.
\newline \newline
For the S24 questionnaire, the app administered and recorded scores using the Erickson S24 scale, ensuring a standardized and objective assessment of speaking confidence. The S24 provides valuable insights into the subjective experience of stuttering, complementing the objective measures of stuttering severity. It consists of 24 statements that evaluate an individual's attitudes and perceptions related to communication and stuttering. Participants respond to each statement with either TRUE or FALSE, indicating their level of agreement. Each statement is associated with a predefined scored option, which represents the opposite of what a typical speaker who does not stutter would choose. If the participant’s response matches the scored response, it is awarded one point. The final S24 score is calculated by summing all points, with higher scores indicating more negative attitudes toward speaking and lower scores reflecting greater confidence and reduced avoidance behaviors.

\section{Results}
A total of 17 participants completed the program. 12 were male, while 5 were female. All of them were between the ages of 18-45, except one male participant aged 15. 5 participants had previously availed speech therapy from a qualified speech language pathologist, 3 others had availed alternative treatment namely hypnosis, homeopathy and coaching from a corporate communications trainer, while 9 had not availed any treatment for their stuttering.
\newline

\begin{table}[ht]
\centering
\renewcommand{\arraystretch}{1.2} % Adjust row height for better readability
\begin{tabularx}{\textwidth}{|X|c|c|c|c|}
\hline
\textbf{Description} & \textbf{Group} & \textbf{Before} & \textbf{After} & \textbf{Improvement} \\ 
\hline
Mean reading score (A) & All & 4.29 & 2.36 & 44.99\% \\
 & Male & 4.10 & 2.00 & 51.22\% \\
 & Female & 4.75 & 3.25 & 31.58\% \\
\hline
Mean speaking score (B) & All & 6.36 & 3.43 & 46.07\% \\
 & Male & 6.00 & 3.00 & 50.00\% \\
 & Female & 7.25 & 4.50 & 37.93\% \\
\hline
Mean duration score (C) & All & 6.29 & 2.71 & 56.92\% \\
 & Male & 5.40 & 2.20 & 59.26\% \\
 & Female & 8.50 & 4.00 & 52.94\% \\
\hline
Mean physical concomitants score (D) & All & 5.29 & 1.93 & 63.52\% \\
 & Male & 4.70 & 1.50 & 68.09\% \\
 & Female & 6.75 & 3.00 & 55.56\% \\
\hline
Mean SSI-4 score (A+B+C+D) & All & 22.21 & 10.50 & 52.72\% \\
 & Male & 20.20 & 8.80 & 56.44\% \\
 & Female & 27.25 & 14.75 & 45.87\% \\
\hline
\end{tabularx}
\caption{\label{tab:ssi4results}Comparison of pre-therapy and post-therapy SSI-4 scores}
\end{table}

\noindent Table \ref{tab:ssi4results} presents the mean pre-therapy and post-therapy scores across different components of the SSI-4 for all participants, as well as for male and female subgroups. Overall, there was a significant reduction in stuttering severity following the intervention, with the mean total SSI-4 score decreasing from 22.21 to 10.50, reflecting a 52.72\% reduction. The greatest improvement was observed in the physical concomitants score, which showed a 63.52\% reduction, followed by the duration score (56.92\%) and speaking score (46.07\%). Male participants demonstrated a slightly higher overall reduction (56.44\%) compared to female participants (45.87\%), with the most pronounced differences seen in reading and speaking scores. These findings indicate that participants experienced meaningful improvements in fluency across multiple dimensions, supporting the effectiveness of the Eloquent app in reducing stuttering severity.
\newline

\begin{table}[ht]
    \centering
    \begin{tabular}{|l|l|c|}
        \hline
        \textbf{Pre-therapy Category} & \textbf{Post-therapy Category} & \textbf{Count} \\
        \hline
        Very severe  & Severe     & 1 \\
        Very severe  & Very mild  & 1 \\
        Severe       & Mild       & 1 \\
        Moderate     & Moderate   & 1 \\
        Moderate     & Mild       & 2 \\
        Moderate     & Very mild  & 2 \\
        Mild         & Very mild  & 5 \\
        Very mild    & Very mild  & 4 \\
        \hline
    \end{tabular}
    \caption{\label{tab:severity}SSI-4 severity category transitions after therapy}
    \label{tab:category_comparison}
\end{table}

\noindent Table \ref{tab:severity} illustrates the changes in stuttering severity categories from pre-therapy to post-therapy based on SSI-4 scores. Notably, one participant showed a dramatic improvement from \textit{Very Severe} to \textit{Very Mild}, while another transitioned from \textit{Very Severe} to \textit{Severe}. Among those initially categorized as \textit{Moderate}, four participants demonstrated improvement, with two reaching \textit{Very Mild} and two moving to \textit{Mild}. All five participants who started in the \textit{Mild} category progressed to \textit{Very Mild}, while four participants who started in \textit{Very Mild} remained in the \textit{Very Mild} category. Importantly, no participant experienced a deterioration in severity level.
\newline

\begin{table}[ht]
    \centering
    \begin{tabular}{|l|c|}
        \hline
        \textbf{Task} & \textbf{Mean Percentage Syllables Stuttered} \\
        \hline
        Pre-therapy reading task & 4.08\% \\
        Pre-therapy speaking task & 8.62\% \\
        Post-therapy reading task & 1.23\% \\
        Post-therapy speaking task & 2.23\% \\
        \hline
    \end{tabular}
    \caption{\label{tab:percentage_syllables} Mean percentage of syllables stuttered pre-therapy and post-therapy}
\end{table}

\noindent Table \ref{tab:percentage_syllables} presents the mean percentage of syllables stuttered before and after therapy during both reading and speaking tasks, along with the relative reduction in stuttering frequency. Pre-therapy, participants stuttered on an average of 4.08\% of syllables while reading and 8.62\% while speaking. Post-therapy, these values decreased to 1.23\% while reading and 2.23\% while speaking, reflecting a relative reduction of 69.9\% and 74.1\%, respectively.
\newline

\begin{table}[ht]
\centering
\begin{tabular}{|l|l|l|l|l|}
\hline
\textbf{Description} & \textbf{Group} & \textbf{Before} & \textbf{After} & \textbf{Improvement} \\ 
\hline
Mean S24 score & All & 21.07 & 14.00 & 33.55\% \\
 & Male & 21.50 & 15.60 & 27.44\% \\
 & Female & 20.00 & 10.00 & 50.00\% \\
\hline
\end{tabular}
\caption{\label{tab:S24}Mean pre-therapy and post-therapy scores for S24}
\end{table}

\noindent Table \ref{tab:S24} presents the mean pre-therapy and post-therapy S24 scores, reflecting changes in participants' self-perceived communication confidence. On average, participants experienced a 33.55\% reduction in their S24 scores, indicating a shift toward more positive speaking attitudes and reduced avoidance behaviors. This improvement suggests that Eloquent not only enhanced fluency but also contributed to increased confidence in real-world communication scenarios.
\newline 

\begin{longtable}{|p{1cm}|p{6cm}|p{1.6cm}|p{1.5cm}|p{1.5cm}|p{1.95cm}|}
\hline
\textbf{S.No.} & \textbf{S24 Question} & \textbf{Scored Option} & \textbf{Before} & \textbf{After} & \textbf{Reduction} \\ 
\hline
\endfirsthead
\hline
\textbf{S.No.} & \textbf{S24 Question} & \textbf{Scored Option} & \textbf{Before} & \textbf{After} & \textbf{Reduction} \\ 
\hline
\endhead

\hline
1  & \raggedright I usually feel that I am making a favourable impression when I talk. & FALSE & 13 & 5  & 61.54\% \\
2  & \raggedright I find it easy to talk with almost anyone. & FALSE & 15 & 10 & 33.33\% \\
3  & \raggedright I find it very easy to look at my audience while speaking to a group. & FALSE & 16 & 10 & 37.50\% \\
4  & \raggedright A person who is my teacher or my boss is hard to talk to. & TRUE  & 14 & 8  & 42.86\% \\
5  & \raggedright Even the idea of giving a talk in public makes me afraid. & TRUE  & 16 & 12 & 25.00\% \\
6  & \raggedright Some words are harder than others for me to say. & TRUE  & 16 & 12 & 25.00\% \\
7  & \raggedright I forget all about myself shortly after I begin to give a speech. & FALSE & 4  & 6  & -50.00\% \\
8  & \raggedright I mix and get along well with people socially. & FALSE & 12 & 9  & 25.00\% \\
9  & \raggedright People sometimes seem uncomfortable when I am talking to them. & TRUE  & 12 & 6  & 50.00\% \\
10 & \raggedright I dislike introducing one person to another. & TRUE  & 10 & 7  & 30.00\% \\
11 & \raggedright I often ask questions in group discussions. & FALSE & 12 & 7  & 41.67\% \\
12 & \raggedright I find it easy to keep control of my voice when speaking. & FALSE & 13 & 9  & 30.77\% \\
13 & \raggedright I do not mind speaking before a group. & FALSE & 13 & 9  & 30.77\% \\
14 & \raggedright I do not talk well enough to do the kind of work I'd really like to do. & TRUE  & 15 & 11 & 26.67\% \\
15 & \raggedright My speaking voice is rather pleasant and easy to listen to. & FALSE & 12 & 7  & 41.67\% \\
16 & \raggedright I am sometimes embarrassed by the way I talk. & TRUE  & 15 & 10 & 33.33\% \\
17 & \raggedright I face most speaking situations with complete confidence. & FALSE & 16 & 6  & 62.50\% \\
18 & \raggedright There are only a few people I can talk with easily. & TRUE  & 16 & 11 & 31.25\% \\
19 & \raggedright I talk better than I write. & FALSE & 15 & 13 & 13.33\% \\
20 & \raggedright I often feel nervous while talking. & TRUE  & 16 & 9  & 43.75\% \\
21 & \raggedright I find it hard to make small talk when I meet new people. & TRUE  & 14 & 11 & 21.43\% \\
22 & \raggedright I feel pretty confident about my speaking ability. & FALSE & 16 & 8  & 50.00\% \\
23 & \raggedright I wish that I could say things as clearly as others do. & TRUE  & 16 & 12 & 25.00\% \\
24 & \raggedright Even though I know the right answer I have often failed to give it because I was afraid to speak out. & TRUE  & 16 & 12 & 25.00\% \\
\hline
\caption{\label{tab:S24_questions}Pre-therapy and post-therapy scored responses for S24 questions}
\end{longtable}

\noindent Table \ref{tab:S24_questions} presents the pre-therapy and post-therapy responses for each item in the S24 questionnaire, highlighting reductions in negative speaking attitudes. The most notable improvements were seen in \textit{I face most speaking situations with complete confidence} (62.50\% reduction in scored responses) and \textit{I usually feel that I am making a favourable impression when I talk} (61.54\% reduction), suggesting a significant boost in self-confidence. Additionally, \textit{People sometimes seem uncomfortable when I am talking to them} and \textit{I feel pretty confident about my speaking ability} both showed a 50.00\% reduction, indicating enhanced comfort in social interactions. Interestingly, one item, \textit{I forget all about myself shortly after I begin to give a speech}, showed an increase in scored responses. This may indicate heightened self-awareness post-intervention, a possible consequence of participants becoming more attuned to their speech patterns. 

\section{Discussion}
The results of this pilot study demonstrate that the Eloquent digital speech therapy program significantly improved both fluency and self-reported speaking confidence in adults who stutter. These findings align with previous research suggesting that structured speech therapy interventions, whether delivered in-person or digitally, can lead to meaningful improvements in stuttering severity \cite{carey2010randomized}\cite{lowe2014review}.
\newline \newline
Beyond fluency gains, participants also reported significant improvements in communication confidence as measured by the S24 scale, suggesting that, in addition to enhancing fluency, Eloquent may help address the psychological and social barriers associated with stuttering, which are critical for long-term success in speech therapy \cite{menzies2008experimental}. The largest improvements were seen in statements related to self-confidence, social ease, and perceptions of how others react to their speech, reinforcing the potential of digital interventions in addressing the psychosocial aspects of stuttering.
\newline \newline
Importantly, no participant experienced a deterioration in fluency or confidence, further supporting the safety and efficacy of the intervention. The transition of several participants from higher stuttering severity categories (e.g., very severe to severe, moderate to mild, and mild to very mild) suggests that Eloquent may facilitate meaningful progress in a relatively short intervention period. However, given the variability in individual responses, further research with a larger and more diverse sample is needed to understand the long-term sustainability of these gains.

\section{Limitations and Future Work}
While this study provides promising evidence for the effectiveness of digital stuttering therapy tools, several limitations must be acknowledged. First, the study sample was relatively small (n=17), which limits the generalizability of the findings. Future studies with larger, more diverse participant groups will help validate these results and provide stronger statistical power.
\newline \newline
Second, the study relied on a short-term intervention period of 15 sessions, with assessments conducted immediately after completion. While significant reductions in stuttering severity and improvements in speaking confidence were observed, the long-term retention of these benefits remains unknown. Future research should incorporate follow-up assessments at multiple time points (e.g. 3 months, 6 months, and 1 year post-therapy) to evaluate the sustainability of fluency gains.
\newline \newline
Third, participant engagement and adherence were self-reported, and while app usage logs were monitored, external factors influencing completion rates were not controlled. Future studies could explore the impact of personalized reminders, gamification, or therapist involvement in improving adherence and engagement.
\newline \newline
Additionally, this study focused primarily on quantitative measures (SSI-4 and S24), which capture speech fluency and self-reported confidence but do not fully reflect qualitative aspects such as real-world communication success, social participation, or emotional well-being. Future work should incorporate qualitative interviews to provide a more comprehensive understanding of user experiences.
\newline \newline
Moreover, this study did not include a control group, which limits the ability to compare the app’s effectiveness against other interventions or a no-treatment condition. Future studies should employ randomized controlled trial (RCT) designs to establish a clearer understanding of Eloquent’s impact relative to standard therapy or placebo conditions.
\newline \newline
Finally, while Eloquent provides structured, self-guided therapy, it does not currently offer real-time feedback or therapist supervision. Hybrid models that integrate AI-driven feedback or periodic therapist check-ins could enhance outcomes by providing more personalized support. Further research is needed to explore how digital therapy can be optimized to complement traditional speech therapy rather than act as a substitute for it.

\section{Conclusion}
This study underscores the potential of digital interventions like Eloquent in transforming stuttering therapy by making treatment more accessible, scalable, and adaptable to individual needs. Traditional speech therapy often faces barriers such as high cost, geographical limitations, therapist availability, low adherence to treatment, and low success in transferring therapy skills to the real world. Digital solutions can bridge these gaps by providing structured, evidence-based interventions that users can access anytime, anywhere. Moreover, advancements in AI and speech recognition can enhance the personalization and effectiveness of such tools, offering real-time feedback and adaptive exercises. However, for digital interventions to be widely adopted, further research is needed to refine their methodologies, assess long-term efficacy, and integrate them into standard clinical practices. By continuing to explore, develop, and validate digital therapy solutions, the field of speech-language pathology can evolve toward more inclusive and innovative approaches that empower individuals who stutter to improve their communication skills with greater ease and confidence.
\newline \newline

\noindent\textbf{\small Declaration of Competing Interests} \newline
Urvisha Shethia and Vedali Inamdar are expert consultants working with Iyaso, which is involved in the development of the digital speech therapy intervention evaluated in this study.

\bibliographystyle{ieeetr}
\nocite{*}
\bibliography{bibliography}

\end{document}